\documentstyle[aps,prl,epsf]{revtex}

\newcommand{\be}{\begin{equation}}
\newcommand{\ee}{\end{equation}}
\newcommand{\bea}{\begin{eqnarray}}
\newcommand{\eea}{\end{eqnarray}}
\newcommand{\bdm}{\begin{displaymath}}
\newcommand{\edm}{\end{displaymath}}

\newcommand{\we}{\wedge}
\newcommand{\Trace}[1]{\mbox{Tr}\{ #1 \}}
\newcommand{\fourg}{\mbox{\boldmath $g$}}
\newcommand{\fourA}{\mbox{$A$}}
\newcommand{\fourast}{\mbox{$\ast$}}
\newcommand{\fourF}{\mbox{$F$}}
\newcommand{\threeg}{\mbox{\boldmath$\overline{g}$}}
\newcommand{\threeA}{\mbox{$\bar{A}$}}

\newcommand{\threeF}{\mbox{$\bar{F}$}}
\newcommand{\bA}{\mbox{\boldmath $A$}}
\newcommand{\bB}{\mbox{\boldmath $B$}}
\newcommand{\bP}{\mbox{\boldmath $P$}}
\newcommand{\bL}{\mbox{\boldmath $L$}}

\begin{document}
\draft
\twocolumn[\hsize\textwidth\columnwidth\hsize\csname @twocolumnfalse\endcsname
\title{Rotating solitons and non-rotating, non-static black holes}
\author{O.\/Brodbeck, M.\/Heusler, N.\/Straumann and M.\/Volkov}
\address{Institute for Theoretical Physics, University of Zurich,
CH--8057 Zurich, Switzerland}
\maketitle

\begin{abstract}
It is shown that the non-Abelian black hole solutions have
stationary generalizations which are parameterized by 
their angular momentum and electric Yang-Mills charge. 
In particular, there exists a non-static class 
of stationary black holes with vanishing angular 
momentum. It is also argued that the particle-like
Bartnik-McKinnon solutions admit slowly rotating, 
globally regular excitations. In agreement with the
non-Abelian version of the staticity theorem, these 
non-static soliton excitations carry electric charge, 
although their non-rotating limit is neutral.
\end{abstract}
]
\narrowtext


\noindent{\bf{Introduction --}}
In recent years it has become obvious that a variety of 
well-known, and rather intuitive, features of  
self-gravitating Maxwell fields are not shared by 
{\it non}-Abelian gauge fields. In particular, and
in contrast to the Abelian situation, self-gravitating 
Yang-Mills (YM) fields can form particle-like configurations
\cite{BK}. Moreover, the Einstein-Yang-Mills (EYM) equations 
also admit black hole solutions which are not uniquely 
characterized by their mass, angular momentum, and YM
charges \cite{BH-HAIR}.
Hence, the celebrated uniqueness theorem for 
electrovac black hole spacetimes \cite{MH-PC} ceases to 
exist for EYM systems. In fact, not even partial results 
of the no-hair theorem can be restored in 
the non-Abelian case \cite{MH-HPA}: 
In addition to the circumstance that spherically symmetric
black holes are, in general, no longer characterized by their 
mass and charges, static black holes need
not even be spherically symmetric \cite{BH-AX}. Moreover, we 
shall show that there exist black hole spacetimes with 
vanishing angular momentum
which are, however, {\it not\/} static.

The new results presented in this letter are based 
on our previous investigations \cite{NS-MV} 
and \cite{OB-MH}. In \cite{NS-MV} we have shown that 
non-Abelian black holes always have rotating counterparts. 
It was also conjectured, that solitons generically 
do not admit rotating excitations. A systematic 
analysis of stationary perturbations revealed that this 
is indeed the case, provided that the EYM system is coupled 
to bosonic matter fields \cite{OB-MH}. 
However, as the {\it pure\/} EYM system 
comprises exclusively massless fields, the polynomial 
fall-off of the background configurations allows for a more 
general asymptotic behavior than the one considered in 
\cite{NS-MV}. Hence, in the {\it absence\/} of bosonic 
fields, one gains an additional degree of freedom, which 
gives rise to the new features described in this letter.

More precisely, we prove the existence of slowly rotating 
Bartnik-McKinnon (BK) solitons \cite{BK}, and establish a
two-parameter family of stationary excitations of the SU(2) 
black hole solutions. In addition to the charged, rotating
solutions found in \cite{NS-MV}, there also exists a
branch of uncharged, rotating black holes, and a branch 
of charged black holes with {\it vanishing\/} angular 
momentum. As these configurations are {\it not\/} static, 
they illustrate that the assumptions entering the non-Abelian 
staticity theorem \cite{SW} are optimal: According to this 
theorem, stationary EYM black hole solutions must be 
static only if they have zero angular momentum 
{\it and\/} vanishing electric YM charge. 
The new solutions demonstrate that the vanishing of 
the electric charge is, in fact, a necessary requirement 
for the configuration to be static. Moreover, the inversion 
of the non-Abelian staticity theorem also predicts that 
rotating excitations of the BK solitons must be charged.

Although it is, by now, mathematically clarified why slow 
rotations of EYM solitons are only possible in the
{\it absence\/} of bosonic fields, we still lack 
a deeper physical understanding of this surprising fact. 
The authors of this letter could not agree on any of 
the heuristic proposals which came up in the discussions.
\\

\noindent{\bf{Stationary perturbations --}}
We start by briefly recalling that the stationary 
perturbations of static EYM configurations are governed by 
a self-adjoint system of equations for a set of gauge 
invariant scalar amplitudes (see \cite{OB-MH} for details). 
A stationary EYM configuration 
(with Killing field $\partial_{t}$, say) is described in 
terms of a stationary metric, $\fourg$, and a stationary 
non-Abelian gauge potential, $\fourA$,
\be
\fourg  \, = \, -\sigma \, (dt + a)^{2}
\, + \, \sigma^{-1} \threeg ,
\label{metric}
\ee
\be
\fourA  \, = \, \phi \, (dt + a) \, + \, \threeA .
\label{potential}
\ee
Here, $\sigma$ and $a = a_{i} dx^{i}$ are a scalar 
field and a one-form on the three-dimensional 
(Riemannian) orbit space with metric $\threeg$, 
respectively,
and so are the Lie algebra 
valued quantities $\phi$ and $\threeA$, describing the 
electric and the magnetic part of the YM field. As we are 
interested in 
{\it perturbations of static, purely magnetic\/}
configurations, both the electric potential and the 
off-diagonal part of the metric vanish for the unperturbed 
solutions, that is, 
$\phi \equiv \delta \phi$ and $a \equiv \delta a$. 

Using the Kaluza-Klein reduction of the EYM action, 
we have shown in \cite{OB-MH} that the non-static 
perturbations, $\delta a$ and $\delta \phi$, decouple 
from the remaining metric and matter perturbations. 
Moreover, in first order perturbation theory, the 
latter do not contribute to the angular momentum. 
The rotational excitations of a static, purely magnetic 
EYM spacetime are, therefore, governed by the linearized 
field equations for the metric perturbation $\delta a$ 
and the electric YM perturbation $\delta \phi$ \cite{NS-MV}.
 
In order to obtain a self-adjoint form of the perturbation 
equations, it is necessary to pass from $\delta a$ to the 
linearized {\it twist potential\/}, $\delta \chi$, defined by
\be
\delta \chi,_{k} = \varepsilon_{kij} \sqrt{\bar{g}}
(\frac{\sigma^{2}}{2} \, d \delta a + 4 \sigma \/
\Trace{\threeF \/ \delta \phi} )^{ij}.
\label{def-twist}
\ee
Here, $\threeF$ is the field strength with respect to 
the magnetic potential $\threeA$, and the spatial indices 
are raised with the $3$-dimensional metric $\threeg$. 
By virtue of this definition, the equations governing 
the non-static, stationary perturbations of the EYM 
system can, eventually, be cast into a formally 
{\it self-adjoint\/} system for the 
{\it gauge invariant scalar\/} quantities $\delta \chi$ 
and $\delta \phi$ \cite{OB-MH}.
(The existence of a generalized twist potential for the 
stationary EYM system follows from the fact that $a$ enters 
the effective action only via the ``field-strength'' $da$; 
see \cite{MH-HPA} or \cite{OB-MH} for details.) 

Since the static background 
solutions under consideration are spherically symmetric,
one can perform a multipole expansion of the 
perturbation amplitudes $\delta \chi$ and $\delta \phi$.
Before doing so, we recall that the background metric, 
$\fourg_{BG} = -\sigma dt^{2} + \sigma^{-1}
\threeg$, is parameterized in standard 
Schwarzschild coordinates by $\sigma(r)$ and $N(r)$,
and the purely magnetic
background gauge potential, $\fourA_{BG} = \threeA$, 
is given in terms of a radial function $w(r)$:
\be
\sigma^{-1} \threeg \, = \, N^{-1} dr^{2} \, + \, 
r^{2} d \Omega^{2} ,
\label{bg-metric}
\ee
\be
\threeA \, = \, (1-w) \, \left( \tau_{\vartheta} 
\sin \! \vartheta \, d \varphi - \tau_{\varphi} \, 
d \vartheta \right),
\label{bg-pot}
\ee
where $\tau_{\vartheta}$, $\tau_{\varphi}$ and 
$\tau_{r}$ are the spherical generators of SU(2), 
normalized such that 
$[\tau_{\vartheta},\tau_{\varphi}] = \tau_{r}$.

The stationary perturbations 
$\delta \chi$ and $\delta \phi$
can now be expanded in terms of spherical ``isospin'' 
harmonics. It turns out that all axisymmetric 
perturbations which give rise to rotational
excitations belong to the sector with 
total angular momentum $j=1$ \cite{NS-MV}. 
The perturbations 
$\delta \chi$ and $\delta \phi$ are 
determined by three scalar amplitudes, 
$\xi_{1}(r)$, $\xi_{2}(r)$ and
$\xi_{3}(r)$ (see \cite{OB-MH} for details),
\be
\delta \chi = 2 \/ \xi_{1} \/ \cos\!\vartheta, \; \; \; 
\delta \phi = \xi_{2} \/ \tau_{r} \/ \cos\!\vartheta -
\frac{\xi_{3}}{\sqrt{2}} \/ \tau_{\vartheta} \/
\sin\!\vartheta .
\label{harmon}
\ee
Using these expansions, the perturbation 
equations finally assume the form of a standard 
Sturm-Liouville equation for the three component real 
vector ${\bf\xi}=(\xi_{1},\xi_{2},\xi_{3})$. One finds
\be
\left( - \frac{d}{dr} \/ r^{2} \bA \/ \frac{d}{dr}
+ \bB  \frac{d}{dr} - \frac{d}{dr} \bB^{T} 
+ \bL + \bP \right) {\bf\xi}  = 0 ,
\label{pert-eq}
\ee
where the $3 \times 3$ matrices $\bA$, $\bB$, $\bL$, 
and $\bP$ are given in terms of the background fields 
$w(r)$, $\sigma(r)$, and $N(r)$. The non-vanishing
matrix elements of $\bA$ and $\bB$ are
\be
\bA = S^{-1} 
\mbox{diag} (-\sigma^{-1},1,1)  , 
\; \;
\bB_{21} = -2 \sigma^{-1} (w^{2}-1) ,
\label{matrix-A-B}
\ee
where we have introduced the metric function
$S$, defined by $S^2 = \sigma / N$.
For the ``angular momentum'' matrix $\bL$ and the
effective potential $\bP$ one finds
\be
\bL  =  \frac{1}{NS}
\left(\begin{array}{ccc}
-2 \sigma^{-1} & 0 & 0 \\
0 & 2(w^{2}+1) & -2 \sqrt{2} \, w \\
0 & -2 \sqrt{2} \, w & (w^{2}+1)
\end{array}\right)  ,
\label{matrix-L}
\ee
\be
\bP  =  -\frac{2}{\sigma}
\left(\begin{array}{ccc}
0 & 0 & \sqrt{2} \, w' \\
0 & 2 S r^{-2} (w^{2}-1)^{2} & 0 \\
\sqrt{2} \, w' & 0 &2 NS w'^{2}
\end{array}\right) ,
\label{matrix-P}
\ee
where a prime denotes differentiation with respect to $r$.
\\

\noindent{\bf{Soliton excitations --}}
The existence of rotational excitations
of the BK soliton solutions is now established as follows:
First, one observes that the Sturm-Liouville 
equation  (\ref{pert-eq}) has {\it regular singular\/} 
points at $r=0$ {\it and\/} $r = \infty$. This is seen by
writing the perturbation equations as a six-dimensional
system of first order equations, and by using the behavior
of the background configurations in the vicinities of the
origin and infinity. 
For instance, one uses
\be
w = 1 -\gamma \frac{2M}{r} + {\cal O}(\frac{1}{r^2})  ,
\; \; 
N = 1 - \frac{2M}{r} + {\cal O}(\frac{1}{r^2})  ,
\label{sr-3}
\ee
and $S = 1 + {\cal O}(r^{-4})$, to conclude that
$r=\infty$ is a regular singular point. This is,
in fact, a peculiarity of the {\it pure\/}
EYM system, for which the {\it polynomial\/} decay of the 
background fields implies that all perturbations are 
{\it massless\/}. (Here, $M$ denotes the total mass and 
$\gamma$ is a parameter characterizing the background 
configuration.) 
Taking advantage of the expansions 
(\ref{sr-3}) shows that the perturbation equations 
decouple in leading {\it and\/} in next-to-leading order. 
In leading order 
one finds a four-dimensional family of asymptotically 
acceptable solutions, behaving like $r^{-\lambda}$,
with $\lambda = 0,\/1,\/2,\/3$. 
Following the standard theory, 
it remains to verify that the fundamental solution 
belonging to $\lambda = 0$ does not exhibit 
logarithmic terms in next-to-leading order. 
In fact, it turns out that
this is the case for all non-negative eigenvalues. Hence, 
one ends up with a {\it four}-dimensional system of 
asymptotically well-behaved local solutions:
\bea
{\bf\xi} & = &  \left(c_0 + \frac{c_1}{r} \right)
\left[{\bf e_1} + {\cal O}(\frac{\ln (r)}{r^2}) \right] +  
\frac{c_2}{r^2} \left[{\bf e_2} + {\cal O}(\frac{1}{r^2}) 
\right]
\nonumber\\ & + &
\frac{c_3}{r^3} \left[ 
\left(1 + (1\!-\!\gamma) \frac{2M}{r} \right) 
{\bf e_3} + {\cal O}(\frac{1}{r^2}) \right] ,
\label{fund-sy-infty}
\eea
where ${\bf e_1} = (0,1,\sqrt{2})$, ${\bf e_2} = (1,0,0)$, and
${\bf e_3} = (0,\sqrt{2},-1)$.
In a similar way one obtains a {\it three}-dimensional
system of admissible solutions in the
vicinity of the origin. Since the BK background solutions
are continuous and regular for $0 < r < \infty$, and since
the perturbation equations are linear, the local solutions 
in the vicinity of $r=0$ and $r=\infty$ admit 
extensions to the semi-open intervals $[0,\infty)$ 
and $(0,\infty]$, respectively.  As the total 
solution-space is six-dimensional, the intersection of 
the regular solution-subspaces is (at least) 
one-dimensional. Hence, all BK soliton solutions 
admit stationary excitations.
\\

\noindent{\bf{Black hole excitations --}}
As for the black hole case, one needs to investigate
the behavior of solutions in the vicinity of the 
horizon, defined by $N(r_H) = 0$. In leading order
the six fundamental solutions behave like 
$(r-r_{H})^{\lambda}$, with $\lambda = 0,\/1,\/2$.
However, a next-to-leading order expansion shows that
two (out of three) solutions belonging to $\lambda = 0$
must be rejected. Since the remaining solutions
are well-behaved, the
subspace of acceptable solutions in the vicinity of 
the horizon is {\it four}-dimensional. Again using the 
regularity of the background configuration for 
$r_{H} < r < \infty$ shows that stationary excitations
of static EYM black holes always exist. However, 
in contrast to the soliton case, the rotating black 
hole configurations are characterized by {\it two\/} 
parameters, rather than only one. Hence, the additional 
degree of freedom at the horizon implies that the 
intersection of the solution subspaces is now
(at least) {\it two}-dimensional.
\\

\noindent{\bf{Discussion --}}
In order to offer an interpretation of the parameters
characterizing the soliton and black hole 
excitations, we consider the local electric YM charge 
and the local Komar angular momentum, defined by flux 
integrals over two-spheres with radius $r$:
\bdm
\tau_z Q(r) = \frac{1}{4 \pi}
\int \fourast \fourF = 
\frac{\tau_z}{3 S} \left[ r^{2} (\xi_{2} + 
\sqrt2 \xi_{3})' + 2 w' \beta \right] ,
\edm
\bdm
J(r) = \frac{1}{16 \pi}
\int \fourast \left(
d \fourg_{\varphi \mu} \we dx^{\mu} \right) = 
- \frac{r^{4}}{6 S} 
\left( \frac{\beta}{r^2} \right)' ,
\edm
where $\beta$ parameterizes the metric perturbation,
$\sigma \delta a \equiv \beta(r) \sin^{2} \! \vartheta d \varphi$
[see Eq. (\ref{metric})].
By virtue of the harmonic expansions (\ref{harmon}) 
and the definition (\ref{def-twist}) of the twist 
potential $\delta \chi$, one obtains an expression for 
$\beta$ in terms of the perturbation amplitudes $\xi_i$, 
\be
\beta \, = \, 2 (w^{2}\!-\!1) \xi_{2} + 
S^{-1} r^{2} \xi_{1}'  .
\label{dual}
\ee
The electric YM charge, $Q$, and the Komar angular 
momentum, $J$, are obtained from the above local 
expressions in the limit $r \! \rightarrow \! \infty$, 
where the asymptotic expansion (\ref{fund-sy-infty}) 
yields $c_{1} =  - Q$ and 
$c_{2} = - ( J + 4 \gamma M c_{0})$.
The leading two terms in the asymptotic 
expansion of the electric potential $\delta \phi$ and 
the metric one-form $\delta a$ are, therefore
(with $q = Q+Mc_{0}(5 \gamma \!-\!3)/2$):
\be
\delta \phi = (c_{0} - \frac{Q}{r}) \/ \tau_z ,
\; \; 
\sigma \/ \delta a =  2 \left( \frac{J}{r} + 
\gamma \/ \frac{4M \/ q}{r^{2}} \right)
\sin^{2} \! \vartheta d\varphi .
\label{asy-1}
\ee

For perturbations of a {\it Schwarzschild\/} background,
the above expressions are, in fact, the exact solutions
of the perturbation equations, where the second term
in $\delta a$ is absent, since $\gamma = 0$ in this case. 
(Note that the Schwarzschild background solution
is given by $w=1$, $S=1$, $\sigma = N=1-2M/r$.)
As $c_{0}$ does not enter the {\it Abelian\/} 
field strength, $F = d \delta \phi \we dt$, 
it has no physical significance and may, 
as usual, be set equal to zero. Hence, as expected,
the stationary excitations of the Schwarzschild solution
are linearized Kerr-Newman solutions, parameterized by 
their charge $Q$ and their angular momentum 
$J$ \cite{NS-MV}. In particular, it is consistent to 
consider perturbations with either $Q=0$ (Kerr) or $J=0$ 
(Reissner-Nordstr\"om).

Returning to the stationary excitations of the non-Abelian 
black holes, we first emphasize that the constant $c_{0}$ 
now has decisive physical consequences. In fact, by
virtue of the covariant derivative, $c_{0}$ enters 
the asymptotic expression for the field strength.
(It does, however, not show up in the expression for 
$Q$, since the corresponding two-form in the formula
for $\fourast \fourF$ is not proportional to the
volume-form of the two-sphere.) As we have argued 
above, one obtains a two-dimensional family of 
excitations in the black hole case, provided that 
the non-trivial asymptotic degree of freedom, $c_{0}$, 
is taken into account. Hence one can, in particular,
consider solutions with either $Q=0$, $J=0$ or,
as in \cite{NS-MV}, $c_{0} = 0$.

We start with the uncharged excitations of EYM 
black holes, $Q=0$. Like in the Abelian case, these
have a non-static metric, $\delta a \neq 0$, and are
rotating, $J \neq 0$. However, despite the fact that 
the electric YM charge vanishes, there now arises a 
non-vanishing electric YM field, 
$E = d \delta \phi + [\threeA, \delta \phi]$. 
Asymptotically, this becomes
\be
E = \tau_z \, \frac{Q}{r^{2}} \, dr  +
2\gamma M\, \frac{c_{0}}{r} \,
\left(\tau_{r} \/ d \cos \! \vartheta  - 
\cos \! \vartheta \/  d \tau_{r} \right) ,
\label{E-asy}
\ee
which vanishes for $Q=0$ only in the Abelian 
case (since then $w=1$, i.e., $\gamma = 0$).
(As already mentioned, the $c_0$ term is tangential 
to the two-sphere and does, therefore, not 
contribute to the electric YM charge. It is also not
hard to verify that the contributions of this term
to the total energy and to the action are finite.)

Even more interesting is the class of stationary 
excitations with $J=0$. Whereas in the Abelian case 
$J=0$ implies $\delta a = 0$, this is no more 
true for perturbations of static EYM black holes: 
Despite the fact that the angular momentum vanishes, 
the perturbed metric is not 
static, as is already seen from the asymptotic 
behavior (\ref{asy-1}). (Again, this effect is 
proportional to $\gamma$, which vanishes for a 
Schwarzschild background.) This shows that there 
do exist EYM black hole solutions with a non-static 
domain of outer communications and vanishing angular 
momentum. It is worthwhile noticing that the local
angular momentum, $J(r)$, does not vanish when 
evaluated for {\it finite\/} values of $r$, in 
particular for $r = r_{H}$; see Fig. 1. 
Hence, these black holes 
have a rotating horizon, $J(r_{H}) \neq 0$, although 
they are non-rotating in the sense that $J=0$. 
(In contrast to this, a Kerr-Newman black hole with
$J=0$ also has $J(r_{H}) =0$, since both quantities 
are proportional to the Kerr rotation parameter.) 
Numerical results for $J(r)$ and $Q(r)$ are shown in Fig. 1. 
We also expect that these black holes have an 
ergosphere (that is, a region in the domain of outer 
communications where the Killing field $\partial_{t}$ 
becomes space-like). This does, however, not show up 
in lowest order perturbation theory, since the metric 
field $\sigma$ is a background quantity within this 
approximation.
\begin{figure}
\epsfxsize=9cm
\centerline{\epsffile{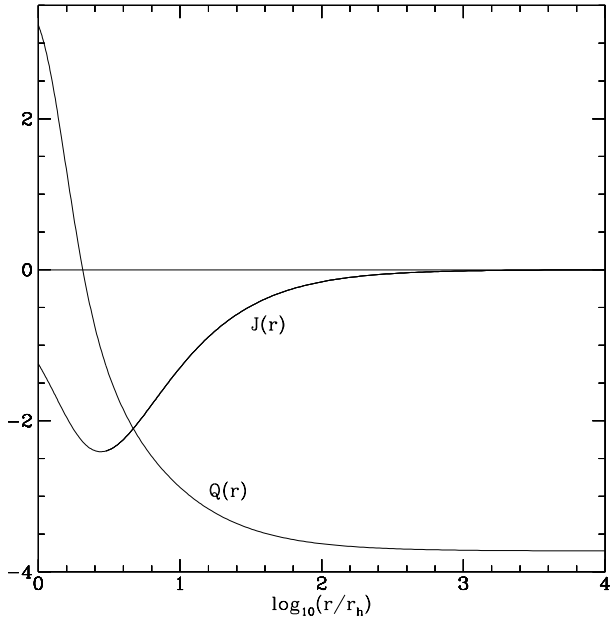}}
\caption{The local charge $Q(r)$ and the local Komar 
angular momentum $J(r)$ for the non-rotating, non-static 
excitation of the non-Abelian background black hole 
with $n=1$, $r_H=1$ .}
\end{figure}

The rotating {\it solitons\/} are characterized by 
one, rather than two parameters. This is due to the 
fact that the solution-space at the origin has one 
dimension less than the solution-space at the horizon. 
Hence, the charge $Q$ and the angular momentum $J$ are 
not independent infinitesimal hair any longer. 
Rotating stationary excitations of the BK solution 
are, therefore, electrically charged.

The Abelian staticity conjecture \cite{Carter}
asserts that stationary, non-rotating black 
hole solutions to the Einstein-Maxwell equations are 
static. In 1992, Sudarsky and Wald were able to
prove this longstanding conjecture and, in addition, 
also established a non-Abelian version of the theorem 
\cite{SW}. Their main result shows that
\be
\Omega_H \/ J - \Trace{\phi_{\infty} \/ Q} \, = 0 
\; \; \Longrightarrow \; \; a \equiv 0 , \; 
\mbox{and} \; E \equiv 0 ,
\label{SW}
\ee
where $\Omega_{H}$ is the angular velocity of the horizon,
$E$ is the electric YM field, and $a$ is the non-static 
part of the metric, defined in Eq. (\ref{metric}).
While this proves that non-rotating, {\it uncharged\/} 
EYM black holes are indeed static, it does not allow the 
same conclusion in the presence of electric YM charges. 
The class of stationary, non-static black
holes discussed above illustrates that $Q=0$ is
not only a sufficient, but indeed a necessary condition
for metric staticity, $a = 0$. Moreover, 
theorem (\ref{SW}) provides an explanation
for the charge-up of rotating {\it solitons}:
Since the first term is not present for soliton 
configurations, one concludes that non-static excitations
($a \neq 0$) must have non-vanishing electric
YM charge. In addition, the theorem also implies that
these solutions can only exist
if $\phi_{\infty}$ does not vanish, which reflects
the crucial importance of the constant term, $c_{0}$, 
in the asymptotic expansion (\ref{fund-sy-infty}).
\\

\noindent{\bf{Conclusions --}}
We have investigated stationary perturbations of static
soliton and black hole solutions to the pure EYM
equations. In contrast to boson stars \cite{Boson-Stars}
or soliton configurations with Higgs fields \cite{OB-MH},
the BK solitons do admit rotating excitations with
continuous angular momentum.
We have argued that this particular feature of the pure
EYM system is due to the slow (polynomial) decay of the
static background configurations. The stationary
excitations of EYM black hole solutions form a 
two-parameter family. In particular, we have presented 
a class of non-static black hole space-times with 
vanishing angular momentum. Both, the existence 
of a second branch of black holes and the charge-up 
of solitons due to rotation are typical non-Abelian 
features of the pure EYM system. While we have shown
earlier \cite{MH-HPA} that the Abelian circularity theorem
does not generalize to EYM systems in a straightforward 
manner, the solutions presented in this letter show that
the same is true for the Abelian staticity theorem: 
In the non-Abelian case, stationary black hole space-times 
with vanishing angular momentum need not be static, 
unless they have vanishing electric YM charges.

\end{document}